\documentclass[conference,compsoc]{IEEEtran}
\usepackage[hyphens]{url}
\usepackage{hyperref}
\hypersetup{breaklinks=true,colorlinks,allcolors=blue}
\usepackage{doi}
\usepackage{longtable}
\usepackage{array}
\usepackage{ragged2e}
\usepackage{comment}
\usepackage{tikz}
\usepackage{pgfplots}
\usetikzlibrary{patterns}
\usepackage{orcidlink}

\ifCLASSOPTIONcompsoc
  \usepackage[nocompress]{cite}
\else
  \usepackage{cite}
\fi

\ifCLASSINFOpdf
  \usepackage{graphicx}
  \usepackage{amsmath}
  
\else
  
\fi

\begin{document}

\title{Beyond Collection: Measuring the Detection Efficacy of \\
Modern Security Logging Standards}

\author{\IEEEauthorblockN{Ryan Holeman\,\orcidlink{0009-0002-8900-1853}}
\IEEEauthorblockA{Beacom College\\Dakota State University\\
Madison, SD, USA\\
ryan.holeman@trojans.dsu.edu}
\and
\IEEEauthorblockN{John Hastings\,\orcidlink{0000-0003-0871-3622}}
\IEEEauthorblockA{Beacom College\\Dakota State University\\
Madison, SD, USA\\
john.hastings@dsu.edu}
\and
\IEEEauthorblockN{Varghese Mathew Vaidyan\,\orcidlink{0000-0002-9737-242X}}
\IEEEauthorblockA{Beacom College\\Dakota State University\\
Madison, SD, USA\\
varghese.vaidyan@dsu.edu}}

\maketitle

\begin{abstract}
Effective security logging is crucial for the timely and accurate detection of cyber threats; however, the relative effectiveness of various industry-standard logging frameworks remains understudied. This paper addresses this critical gap by presenting the first systematic evaluation of modern security logging standards utilizing a novel methodology built upon the automated Security Exploit Telemetry Collection (SETC) framework. SETC systematically generates reproducible exploit scenarios in containerized environments, collecting rich telemetry across multiple logging standards, including CIM (Common Information Model), OCSF (Open Cybersecurity Schema Framework), and ECS (Elastic Common Schema). The detection efficacy of each logging standard is quantified by measuring telemetry completeness and exploit detectability across standardized logs through detailed experiments involving 50 diverse remote code execution vulnerabilities. The resulting findings identify critical gaps and reveal significant differences in logging standards' abilities to capture key attack indicators. Our contributions include a novel evaluation methodology that enables scalable and reproducible analysis of exploit telemetry, as well as new findings that provide clear, evidence-based guidance for security practitioners to make informed decisions about adopting logging standards.
\end{abstract}

\IEEEpeerreviewmaketitle

\section{Introduction}
As emerging software vulnerabilities continue to threaten enterprises, software, and services\cite{verizon2025dbir}, there is a critical need to improve security telemetry and logs in order to detect and respond to vulnerability exploitation. In order to structure this telemetry, security practitioners and organizations often deploy logging standards to guide what infrastructure telemetry should be recorded and the format it should follow. The templates provided by logging standards provide uniform schemas across diverse data sources and promote consistent analysis, detection, and interoperability across security tools and platforms. Some of the most widely adopted structured log formats include Splunk's Common Information Model (CIM)\cite{splunk}, Elastic's Elastic Common Schema (ECS)\cite{ecs}, and the Open Cybersecurity Schema Framework (OCSF)\cite{ocsf}.

While logging standards guide what data should be logged and how it should be organized, there is a lack of information regarding their effectiveness in identifying various types of exploitation or malicious activity. It remains unknown whether specific standards provide better exploit detection than others. There is a crucial need for quantitative analysis to determine which logging models are most effective at capturing telemetry that enables security teams to detect attacks and exploitation attempts.

This paper introduces a novel methodology to evaluate and measure the effectiveness of various logging standards through a design science approach with applied experimentation. The design science component of the research incorporates feature contributions to the Security Exploit Telemetry Collection (SETC)\cite{SETC} framework. The SETC framework is an open-source, automated framework that can generate reproducible vulnerability exploit data at scale for rich security research.

Using SETC, we generate and analyze activity telemetry for 50 remote code execution vulnerability exploits. The telemetry is formatted in various logging schemas for attacks and exploitation in realistic environments. This data allows us to perform an applied experimentation evaluation to analyze three critical questions: what gaps exist in each logging model analyzed for exploitation detection, how do logging models compare to each other in detection telemetry capability, and what metric can be used for measuring the effectiveness of logging standards in a security context?

The motivation for this research is threefold. First, there is a lack of empirical data on logging standard effectiveness for security telemetry. Second, there is insufficient tooling to support the analysis of logging standard effectiveness for exploitation discovery. Finally, security practitioners need quantitative evidence to make informed decisions about which logging standards to adopt.

To address these gaps, this research makes the following primary contributions:
\begin{enumerate}
    \item Presents a reproducible methodology leveraging the Security Exploit Telemetry Collection (SETC) framework for measuring how well security logging standards support detection across an end-to-end exploit chain.
    \item Defines two quantitative metrics that capture preservation of detection-relevant signals through normalization and attack-chain detection coverage using standardized logs.
    \item Empirically compares CIM, OCSF, and ECS on a benchmark of 50 RCE vulnerabilities executed in controlled, repeatable environments.
    \item Identifies practical logging blind spots that limit detectability and highlights which telemetry sources most consistently enable detection.
\end{enumerate}

The remainder of this paper is organized as follows. Section 2 provides a background of logging standards and SETC. Section 3.1 provides an overview of the scope of the evaluation, followed by the evaluation approach in Section 3.2. Section 3.3 presents the results from our assessment of logging standards. Section 4 discusses the limitations of the research. Section 5 reviews work related to the research. Section 6 presents ideas for future work on this research and proposed future contribution work for the SETC framework. Finally, in the conclusion of Section 7, we discuss the implications of our findings and provide recommendations for security practitioners.

\section{Background}
This section provides foundational context for understanding the evaluation of security logging standards and their practical implementation. We first detail the Security Exploit Telemetry Collection (SETC) framework, a specialized platform for generating reproducible vulnerability exploit data at scale within containerized environments. SETC's architecture enables consistent security telemetry collection across multiple standards, making it an ideal tool for comparative analysis.

We then examine prominent logging standards in the security industry, including the CIM, OCSF, ECS, Common Event Format (CEF)\cite{cef}, and Unified Data Model (UDM)\cite{udm}. Each standard presents unique characteristics, structures, field naming conventions, and extensibility. Each logging standard also has various community adoptions based on age, tooling support, and various other factors. These differences have significant implications impacting coverage and detection capabilities, particularly when monitoring active vulnerability exploitation of targets for various vulnerability classes.

\subsection{SETC}

The SETC framework provides automated workflows and generates reproducible vulnerability exploit data for security research. Through the use of configurable container environments and templated workflows, SETC can execute end-to-end attacks on vulnerable systems and record rich telemetry during vulnerability exploitation. The framework addresses key limitations commonly found in other systems that provide attack data in security research. 

The SETC framework addresses fundamental limitations in defensive security research by providing an automated, systematic approach to vulnerability exploitation data generation that overcomes the constraints of existing methods. Many research approaches currently rely on manual vulnerability exploitation and data collection processes. This approach is laborious, error-prone, and difficult to scale. Other techniques, such as utilizing real-world incident data, are constrained by limited access to specific data sources and cannot reproduce attack conditions for comprehensive analysis. In contrast, SETC's containerized architecture enables automated end-to-end exploitation across diverse vulnerability attack scenarios while simultaneously collecting telemetry from multiple sources. 

\subsubsection{Architecture Overview}

SETC is orchestrated by a Python application that utilizes container technologies to manage the entire exploit lifecycle. Within SETC, the framework runner serves as the core component, automatically generating multi-container environments based on JSON configuration files. These configuration files specify which exploits to run, what telemetry to collect, and which logging standards to use. The framework utilizes three types of containers within the environment to perform exploitation and produce telemetry data. These container types include:

\begin{enumerate}
\item \textbf{Vulnerable Systems} host specific vulnerabilities that are fully configured and initiated upon start, sourced from repositories such as Vulhub \cite{vulhub} and proof-of-concept collections. 

\item \textbf{Attacker Systems} contain exploitation tools and scripts capable of compromising vulnerable services, supporting both offensive security frameworks, such as Metasploit \cite{metasploit}, and custom exploit scripts. 

\item \textbf{Auxiliary Systems} provide helper services, including telemetry collection, log pipelines, and data storage. Most of the telemetry collection containers utilize sidecar \cite{sidecar} patterns to passively monitor and collect data from the containers during system exploitation.
\end{enumerate}

\subsubsection{Telemetry Collection}

SETC's telemetry collection capabilities encompass both raw security event data and standardized logging formats, enabling comparisons that assess the effectiveness of logging standards. This approach helps identify which security-relevant signatures are preserved or lost during the conversion of log data. By quantifying vulnerability exploitation telemetry across various logging formats, SETC provides key empirical data for evaluating how well each standard captures the necessary telemetry for detecting active exploitation. The framework's automated and reproducible approach ensures that, as logging standards change or new vulnerabilities emerge, this telemetry can be updated or completely regenerated for future analysis.

\subsection{Logging Standards}
\begin{figure}[ht!]
    \centering
    \includegraphics[width=1.0 \linewidth]{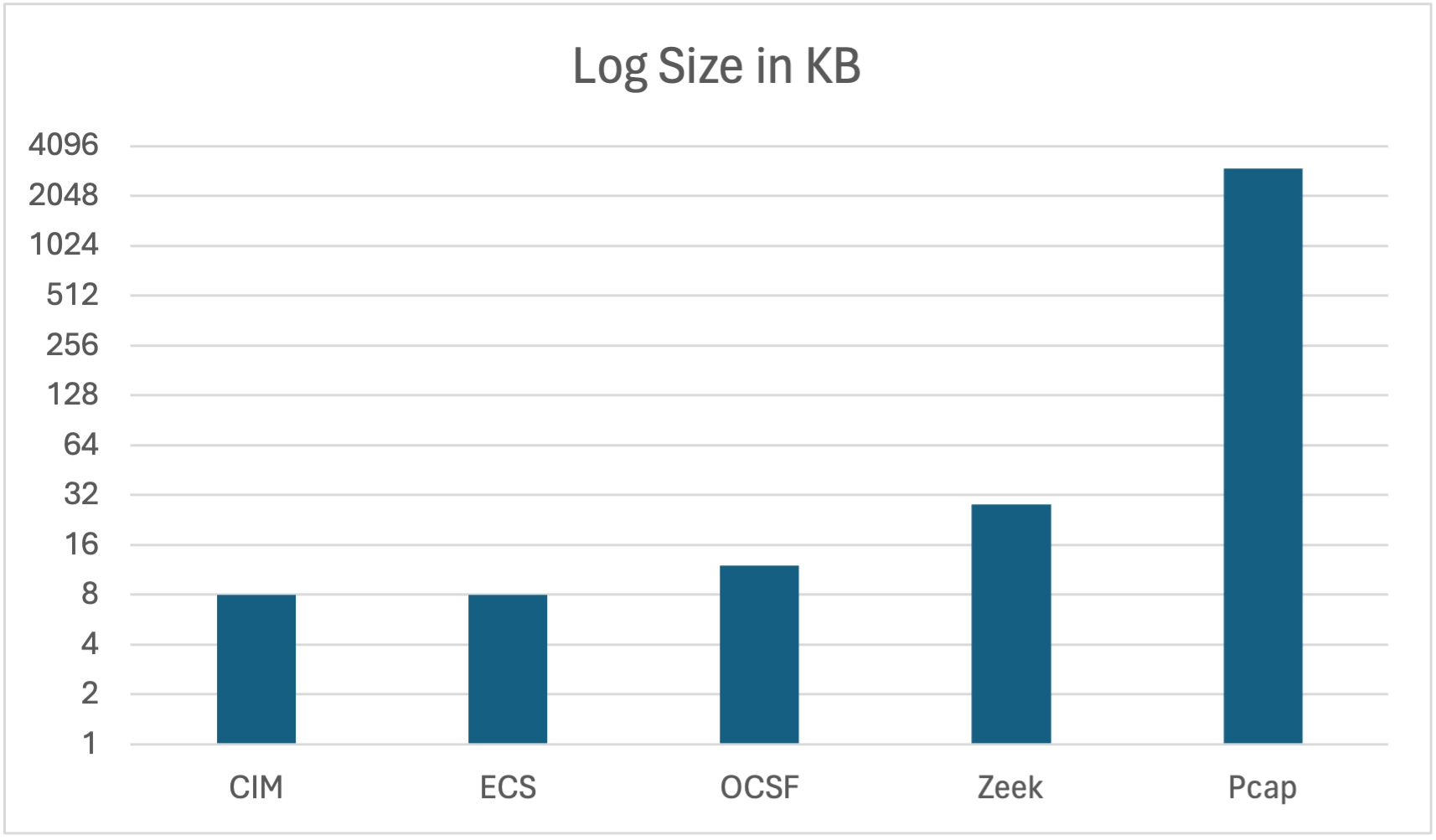}
    \caption{SETC network and HTTP log sizes for CVE-2024-38856}
    \label{fig:log_size}
\end{figure}

Standardized logging formats provide templates specifying what fields should be logged for specific event types. As seen in Figure \ref{fig:log_size}, log standards can offer storage optimizations and query acceleration; however, their primary value for security practitioners lies in defining telemetry scope and format. This scope and format dictate what data sources should be collected and how they can be interpreted by detection rules and analyzed during security investigations.

\subsubsection{Common Information Model (CIM)}
Splunk's Common Information Model (CIM) is a flexible logging format designed to provide uniform field formatting for extraction, grouping, tool interoperability, and query acceleration. CIM's underlying format, event types, and required fields are defined in JSON templates for each supported log type. Logs conforming to CIM must include defined required fields for each event type, but the standard provides flexibility to incorporate supplementary fields beyond those specified in each template. As CIM is implemented via JSON specification, it can be utilized by any tool or system. The format is not limited to Splunk's SIEM ecosystem. With Splunk's widespread adoption as a SIEM platform, numerous third-party services and tools now provide CIM-formatted logs.

\subsubsection{Open Cybersecurity Schema Framework (OCSF)}
The Open Cybersecurity Schema Framework (OCSF) represents one of the newest and most comprehensive logging formats. Developed as a collaborative effort among major security vendors including Splunk, AWS, Broadcom, Salesforce, Rapid7, Tanium, Cloudflare, Palo Alto Networks, DTEX, CrowdStrike, IBM Security, JupiterOne, Zscaler, Sumo Logic, IronNet, Securonix, and Trend Micro, OCSF distinguishes itself as an open-source standard enabling community collaboration and contributions. Following the pattern of other modern formats, OCSF implements a JSON schema with flexible field and metadata extensions beyond its core definitions.

\subsubsection{Elastic Common Schema (ECS)}
Released in 2018, the Elastic Common Schema (ECS) provides standardized field definitions for Elasticsearch implementations to facilitate correlation across various use cases spanning IT operations and security. In 2023, a significant collaboration between Elastic and OpenTelemetry merged OpenTelemetry's Semantic Conventions into ECS, creating benefits for both event producers and consumers. As Elastic has gained popularity as an enterprise logging and monitoring solution, ECS adoption has expanded considerably, resulting in a growing ecosystem of open-source cybersecurity detection rules based on ECS-formatted data.

\subsubsection{Common Event Format (CEF)}
The Common Event Format (CEF), created by ArcSight (now part of Micro Focus), is one of the most widely adopted and oldest logging formats, having been proposed in 2009. CEF's longevity has contributed to its widespread enterprise adoption and frequent use in academic research. The format utilizes a simplistic flat hierarchy of key-value pairs for various log types. However, this age is also one of the format's primary limitations. Due to lack of modernization, many of CEF's supported log types and fields provide less visibility into contemporary protocols, services, and attack vectors that have emerged since its inception.

\subsubsection{Unified Data Model (UDM)}
Google's Unified Data Model (UDM) was created to ensure logs can be seamlessly utilized across Google's security portfolio, including Security Command Center, Siemplify, and Chronicle. While UDM adoption is typically concentrated in enterprises heavily invested in Google's ecosystem for hosting, logging, and security services, it can be implemented with non-Google tooling. Structurally, UDM follows hierarchical logging patterns similar to other standards, with logs typically serialized into JSON or Protobuf formats for compatibility with various analysis tools.

\subsubsection{Dynamic Logging Templates}
Beyond these established standards, emerging research explores dynamic logging template generation. For example, projects like Log2Vec\cite{log2vec} can generate logging templates using natural language processing and machine learning techniques. While these approaches show promise in academic research, particularly for documentation, indexing, and normalizing unstructured logs, they generally lack the benefits of industry-adopted standards, such as pre-existing detection rules and cross-tool interoperability.

\begin{table*}[ht]
\centering
\caption{Logging Standards}
\label{tab:log_standards}
\begin{tabular}{ |p{2cm}||p{6cm}|p{2.5cm}|p{3cm}|  }
 \hline
 \multicolumn{4}{|c|}{Logging Standards} \\
 \hline
 Abbreviation& Name &Associated Organization&SETC Support\\
 \hline
 CIM &   Common Information Model  & Splunk   & True\\
 OCSF & Open Cybersecurity Schema Framework & Various &  True\\
 ECS &   Elastic Common Schema  & Elastic & True\\
 CEF  & Common Event Format    & ArcSight &   Partial\\
 UDM    &Unified Data Model & Google &  Partial\\
 
 \hline
\end{tabular}
\end{table*}

\subsubsection{Comparative Analysis}
The effectiveness of these logging standards in capturing security-relevant events has significant implications for detection capabilities. For instance, if a standard's HTTP template provides fields for cookie data while another omits them, this directly impacts the ability to detect exploits targeting cookie parsing vulnerabilities. Our research utilizes the SETC framework to quantitatively evaluate how these differences affect real-world detection capabilities across various attack scenarios and vulnerability classes. All log standards considered for this research are shown in Table \ref{tab:log_standards}.

\section{Evaluation}
This section presents our systematic approach to evaluating logging standard effectiveness in a security context. We developed a comprehensive methodology to quantitatively measure how well various logging standards capture the telemetry necessary for detecting exploitation of security vulnerabilities. Our evaluation combines controlled experimentation using the SETC framework with detailed analysis of detection capabilities for real-world exploit scenarios.
The evaluation addresses three core research questions:
\begin{enumerate}
    \item What gaps exist in each logging model for exploitation detection?
    \item How do logging models compare to each other in detection telemetry capability?
    \item What metrics can most effectively measure logging standard efficacy in a security context?
\end{enumerate}
Our approach consists of three interconnected components: a carefully defined evaluation scope, a methodology for measuring detection capability, and a quantitative analysis of results across multiple dimensions. The following sections present details on each component. 

\subsection{Evaluation Scope}
One of the most significant limiting factors of vulnerability research is the availability of vulnerable systems and their associated vulnerability exploits \cite{limitedexploits}. While the SETC framework aims to provide researchers with a large corpus of vulnerable services and associated exploits, creating and curating these populations is a slow and manual process. Each vulnerability included in our evaluation requires identifying or building functional containerized environments hosting the vulnerable service, locating reliable exploits that successfully compromise the vulnerability, extensive testing to ensure exploit reliability and reproducibility, and integration work to incorporate both vulnerable systems and exploits into the SETC framework architecture.

This multi-step process for each vulnerable service and associated exploit creates substantial overhead that limits the scope of our evaluation. Despite these constraints, our population selection approach ensures that each included vulnerability represents a high-quality, reproducible test case that enables meaningful comparison across logging standards. This section outlines the specific parameters that define our evaluation boundaries given these practical limitations.

\subsubsection{Attack Graph and Methodology}
\begin{figure*}[t!]
    \centering
    \includegraphics[width=1\linewidth]{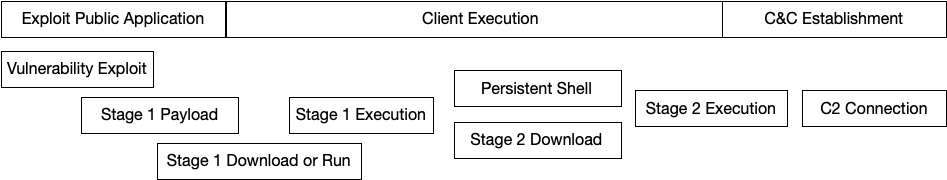}
    \caption{Vulnerability population attack graph and associated exploit activity}
    \label{fig:attack-graph}
\end{figure*}

Our research population for this paper focuses exclusively on remote code execution (RCE) vulnerabilities. RCEs were chosen as a focus because they represent one of the most prominent and severe attack vectors in modern enterprise environments. Vulnerabilities selected for this research follow a standard three-phase attack chain. This chain is aligned with the MITRE ATT\&CK framework\cite{mitre}, beginning with Initial Access through Exploit Public-Facing Application tactics, progressing to Execution via Exploitation for Client Execution techniques, and concluding with Command and Control establishment through Application Layer Protocol methods. This attack chain is illustrated in Figure \ref{fig:attack-graph}.

The initial access phase involves exploiting vulnerable services exposed to a network, typically through HTTP-based attack vectors targeting web applications or open service ports. Following a successful compromise, the execution phase encompasses post-exploitation command execution on the compromised system. This command execution involves process creation and system command invocation to establish attacker control. The final phase focuses on establishing persistent communication channels with reverse TCP connections through remote access tools or commands to maintain persistent access.

This attack progression represents a very common exploitation pathway observed in real-world incidents. This helps ensure our evaluation captures log telemetry across multiple sources during the complete attack lifecycle rather than focusing solely on initial compromise indicators. Using this attack flow across all test cases also ensures we maintain consistent evaluation criteria while reflecting authentic adversarial behavior patterns.

\subsubsection{SETC Compatibility Requirements}

The vulnerability selection for this research is constrained by the current feature capabilities of the SETC framework. These feature constraints impose specific requirements on the types of vulnerabilities that can be  included in the evaluation. As of this writing, the framework currently supports only Linux-based systems for hosting vulnerabilities and performing exploitation. These constraints stem from the framework's container-based architecture. This operating system constraint, while limiting to our scope, ensures consistency across all test environments. It also aligns with the predominant target platform for remote exploitation attacks due to the prevalence of Linux systems in server environments where RCE vulnerabilities are most commonly exploited. 

Another current feature constraint present in the SETC framework is that all vulnerabilities in our corpus must have corresponding exploit modules within the Metasploit Framework. SETC can support exploit modules beyond Metasploit, though Metasploit modules currently have the most comprehensive integration and provide faster development capabilities. Therefore, all vulnerabilities in our corpus have corresponding exploit modules within the Metasploit Framework. This approach eliminates variability introduced by custom exploit scripts and ensures consistent attack methodology across all test cases. 

The final population constraint imposed by the use of SETC is that all vulnerable systems must be deployable using Docker or Docker Compose configurations. While this limitation does constrain population selection, it is less of a constraint %
because the container design of SETC facilitates rapid deployment and teardown of test environments while maintaining strict isolation between experiments. This reproducible execution environment is essential for the systematic telemetry collection done in the evaluation.

\subsubsection{Vulnerability Population Sampling}
Our sampling methodology combines purposive and convenience sampling strategies to create a representative vulnerability corpus. These sampling methods are required to:
\begin{itemize}
    \item Align vulnerabilities with our defined attack graph.
    \item Conform to the limitations of SETC.
    \item Address the limited availability of vulnerable service containers.
    \item Address the limited availability of reliable vulnerability exploits.
    \item Address time and resource constraints associated with testing and incorporating vulnerabilities and service containers into the SETC framework.
\end{itemize}

This research uses purposive sampling to select vulnerabilities that align with our defined attack graph and meet SETC's technical compatibility requirements. The core criteria for purposive sampling included selecting Linux-based RCEs with exploits available in Metasploit. It was also essential to test and verify that these exploits could deploy remote access tools or use living-off-the-land\cite{layoftheland} techniques to connect to a command and control server, as command and control was a component of our proposed attack graph. 

Convenience sampling primarily addresses the time and resource challenges of vulnerability corpus construction. Prioritizing the selection of vulnerabilities with readily available containerized vulnerable systems from established repositories such as Vulhub and other open-source security research projects greatly reduced the time and effort of creating a population set. This pragmatic approach balances research rigor with practical feasibility. It is a realistic approach accepting the fact that the effort required to create custom vulnerable environments for every potential vulnerability would be prohibitive. The combination of these sampling strategies ensures the vulnerabilities in the corpus are both methodologically sound and practically implementable.

\subsubsection{Logging Standards Selection}
Our evaluation includes only logging standards that are fully implemented within SETC and support comprehensive telemetry collection across our attack methodology. This evaluation set consists of three logging standards. The CIM standard provides complete implementation supporting network, HTTP, and process logging templates, making it suitable for comprehensive evaluation across all phases of our attack graph. The OCSF standard offers full compatibility with required event classes for our attack methodology, representing the newest generation of collaborative logging standards. The ECS standard provides comprehensive support for all necessary logging categories and has gained significant adoption in enterprise environments.

For this research, we excluded CEF and UDM from our evaluation due to incomplete implementation in the current version of SETC. While these standards have significant industry adoption, their exclusion ensures fair comparison across logging standards with equivalent implementation completeness, adoption, and telemetry coverage. The exclusion of these standards prioritizes evaluation quality over breadth. 

\subsubsection{Population Summary}
The final population set for this evaluation consists of 50 vulnerabilities. Although CVE descriptions and records do not explicitly categorize vulnerabilities by protocol or attack type such as RCE, our analysis of available Metasploit modules and containerized vulnerable systems revealed a natural clustering around HTTP-based services. The vulnerable service distribution includes 40 HTTP-based web service vulnerabilities, representing 80\% of our corpus, with the remaining 10 vulnerabilities consisting of various network service vulnerabilities (20\%). This distribution reflects the predominance of web-based attack vectors in contemporary threat landscapes documented in recent threat intelligence reports\cite{verizon2025dbir}, while ensuring sufficient representation of non-HTTP attack vectors to maintain corpus diversity.

All vulnerabilities in our corpus are confirmed remote code execution vulnerabilities with verified Metasploit Framework compatibility, ensuring consistent exploitation methodology across all test cases. Each vulnerability follows our defined three-phase attack graph, providing uniform evaluation criteria while maintaining realistic attack progression patterns. The entire corpus is deployable in containerized environments using SETC. This provides reproducible, isolated testing essential for the logging standard evaluation and comparison.

\subsection{Evaluation Approach}
The data for this evaluation is generated with SETC using a custom-created configuration file that includes the 50 exploits and vulnerable services selected for our population. For this data generation, SETC (version 1.0) was configured to generate and store data in the following formats: raw telemetry, CIM (version 6.0.3), OCSF (version 1.6.0), and ECS (version 9.0.0). SETC was also configured to store the generated data in both a Splunk SIEM instance and flat file storage. The configuration file and logs generated for this research are fully open-source. Information on where to access these data sets can be found in the final Availability section of the paper.

The evaluation process for the data in this research uses a six-step approach that measures logging standard effectiveness in detecting exploitation across our defined attack graph described in Section 3.1.1. This approach uses both comprehensive manual analysis and large language model (LLM) analysis to quantify detection capabilities and identify gaps in standardized logging formats. As the vulnerabilities and exploits were selected to adhere to the defined attack graph, the study can be limited to three core logging standard sources: network, process, and web/HTTP logs.

\subsubsection{Log Source Analysis}
We examine three log sources for each vulnerability in our population for each in-scope logging standard. Each log source corresponds to different phases of our defined attack chain. The sources include network, process,  and web logs.  

Network logs capture network-level communication between attack and target systems. Network logs in standardized formats rarely contain exploit signatures due to the fact that they are highly aggregated and condensed similar to NetFlow\cite{netflow} logs. While these highly condensed network logs typically have limited standalone detection value for RCE vulnerabilities\cite{netflowlimitations}, they serve as crucial validation sources for both ingress exploitation traffic, egress command-and-control connections, and data exfiltration.  In our analysis these logs are used to validate that successful connections were made to a vulnerable host and that command and control connections occurred.

Process logs represent a high-impact telemetry source for our evaluation. These logs capture the creation of new processes resulting from successful exploitation, providing direct evidence of code execution on compromised hosts. Depending on the logging standard's implementation, process logs may contain detailed information about exploit execution, indicators of compromise (IOCs), attack Tactics, Techniques, and Procedures (TTPs) and details of post-exploitation command execution. The comprehensiveness of process logging fields directly correlates with detection effectiveness for the attack model used in this research.

As  80\% of our population corpus contains HTTP-based vulnerabilities, web logs provide a primary source of exploit signatures for web-based exploits. These logs can capture HTTP payloads, vulnerable web service interactions, and exploitation attempts through URL paths, headers, and HTTP arguments. For web service vulnerabilities, web logs serve as a primary detection point in our attack chain and often contain the most definitive sources of exploitation signatures in our environment.

\subsubsection{Data Preparation and Analysis Process}
Once data has been collected, a manual and automated analysis of the data produced by SETC must be performed to measure and gauge the effectiveness of each log standard analyzed in this study. This analysis is separated into six phases that systematically evaluate detection capabilities across our defined attack graph.

\textbf{Phase 1: CVE Classification and Tagging -} All raw telemetry and standardized logs are tagged with their corresponding CVE identifiers. This classification process occurs across all logs generated by SETC for the successful exploitation of all vulnerabilities defined in the population set. The correlation of logs to CVEs is guided by SETC attack host logs, which provide hostnames and timestamps correlating to start, end, and duration times for each CVE. This temporal correlation ensures accurate attribution of log events to specific exploits, establishing a foundation for systematic analysis across all subsequent phases.

\textbf{Phase 2: Exploit and Payload Analysis - } A manual analysis of each vulnerability and exploit is performed to establish expected signatures. During this phase, the source code for each exploit is analyzed to identify what the exploit code does, what signatures exist, and what protocols it is sent over. This helps identify if false negatives exist in the raw logs or the log standard logs for the initial access exploit stage of the attack. For each exploit, the second-stage C2 payload is also inspected to identify the type of process and protocol deployed to the target system. This analysis helps identify if false negatives exist in the raw logs or the log standard logs for the execution and C2 connection stage of the attack.

\textbf{Phase 3: Raw and Standardized Telemetry Analysis - } We conduct a comprehensive manual analysis of all raw and standardized telemetry logs classified in Phase 1. The goal of this phase is to identify if key high-fidelity attack signatures exist for each phase of our attack graph. For initial access through exploit public-facing application tactics, we look for network signatures specific to the vulnerability exploit. For execution via exploitation, we look for signatures related to either the download of the second-stage payload or process execution signatures related to the second-stage payload. Finally, for command and control establishment, we look for process or network connection signatures to the C2 server.

\textbf{Phase 4: LLM-Assisted Signature Identification - }In this phase, we utilize LLMs, specifically Claude Opus 4, to identify and classify all low-fidelity detection signatures that exist in raw and standardized logs. Each attack has many of these types of signatures that, on their own, do not provide concrete evidence that an attack has occurred. These signatures include IOCs or TTPs such as user agents, endpoint enumeration, connection characteristics, or process behaviors. A uniform report is produced and stored for each LLM analysis, allowing for later metric analysis and comprehensive coverage of potential detection indicators.

\textbf{Phase 5: Validation and Error Correction - } It was found that using LLMs with RAG context produces numerous errors and hallucinations for this type of analysis. Before using the reports from Phase 4, each report must be thoroughly reviewed through manual analysis. When errors or hallucinations are identified, they are tracked, added to the appendix of each report, and finally, a new report is produced. This process is then repeated for each newly produced report until accuracy is confirmed. This validation step ensures the reliability of our automated analysis while maintaining the efficiency benefits of LLM assistance.

\textbf{Phase 6: Gap Analysis and Metric Calculation - }The final phase involves a comparative analysis of the signatures identified in Phases 3 and 4 against all logging standards utilized in this research. Gaps are identified where exploit signatures exist in raw logs but are absent from standardized formats. This analysis considers both missing data fields and insufficient field granularity. The results of this phase are then used to calculate effectiveness and detection scores, which quantify the telemetry preservation and detection capabilities of each logging standard across our vulnerability corpus.

\subsubsection{Measurement Methodology}

Our evaluation employs two complementary metrics to quantify the effectiveness of logging standards in capturing security-relevant telemetry. These metrics provide both granular and holistic perspectives on how well each standard preserves the information necessary for detecting vulnerability exploitation across our defined attack graph.

The effectiveness score represents a quantitative measurement of telemetry preservation throughout the standardization process. This metric calculates the percentage of exploit signatures identified in raw telemetry logs that remain detectable after transformation into standardized logging formats. For each vulnerability in our corpus, we systematically enumerate all security-relevant signatures present in the raw network captures, system audit logs, and application logs. We then identify which of these signatures persist in the corresponding standardized log entries. The effectiveness score is calculated as the ratio of signatures preserved in standardized logs to the total signatures present in raw telemetry, expressed as a percentage. This approach quantifies the information loss that occurs when comprehensive raw telemetry is processed into condensed standardized formats, providing insight into how logging standards balance completeness with practicality.

{\small
\begin{equation}
EffectivenessScore =
\frac{SigsInStandardLogs}{SigsInRawLogs} \times 100
\label{eq:eff_score}
  \end{equation}
  }
The cardinal detection score complements the effectiveness metric by evaluating whether standardized logs contain sufficient information to identify complete attack chains. Rather than measuring individual signature preservation, this binary metric assesses whether logs formatted according to each standard provide adequate visibility of high-fidelity detection signatures across all three phases of our defined attack methodology. For a vulnerability to receive a positive detection score, the standardized logs must contain clear evidence of initial access through exploit signatures in HTTP or network logs, execution through process creation or command execution indicators, and command and control establishment through outbound connection evidence. This holistic measurement reflects real-world security monitoring scenarios where analysts rely exclusively on standardized logs for threat detection.

{\small
\begin{equation}
Detection Rate = \frac{VulnsWithCompVis}{TotalVulns} \times 100
\label{eq:det_rate}
\end{equation}
}

The combination of these metrics enables a comprehensive evaluation of logging standard performance. While the effectiveness score reveals granular gaps in telemetry coverage, the cardinal detection score determines practical utility for security operations. Together, they provide both detailed technical assessment and operational relevance, ensuring our evaluation addresses the needs of both security researchers seeking to improve logging standards and practitioners selecting standards for deployment in production environments.

\subsubsection{Evaluation Example}
To illustrate the six-phase evaluation process, we examine the analysis process for CVE-2014-6271, commonly known as Shellshock. This vulnerability serves as an ideal case due to its well-understood exploit mechanism and a clear attack progression that follows our defined three-phase attack graph. The Shellshock vulnerability utilizes remote code execution through malicious environment variable injection in bash, making it particularly suitable for demonstrating how our evaluation captures telemetry across multiple logging sources.

\textbf{CVE Classification and Tagging:} The evaluation begins with the organized classification of all telemetry generated during the Shellshock exploitation. SETC automatically segregates logs by vulnerable host, attack host, and network activity into distinct directories. Each log entry, including raw packet captures, process telemetry, and standardized format logs, receives CVE correlation tags based on SETC's attack orchestration. For this vulnerability, logs span from the initial HTTP request targeting /victim.cgi through successful command and control establishment on port 4444, enabling precise attribution of all events to the Shellshock exploitation chain.

\textbf{Exploit and Payload Analysis: } Manual inspection of the Metasploit module reveals the exploit's architecture. The initial exploit embeds malicious bash commands within HTTP headers, specifically crafted to trigger the environment variable parsing vulnerability. The exploit constructs an exploit payload containing bash commands within the User-Agent header, which, when processed by vulnerable CGI scripts, executes the arbitrary commands. The exploit's second-stage payload deploys a reverse shell that creates an executable in the /tmp/ directory with a random five-character name and establishes a connection back to the attacker on port 4444.

\begin{figure}[h!]
    \centering
    \includegraphics[width=1\linewidth]{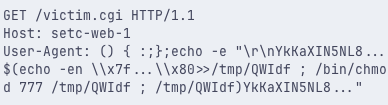}
    \caption{Raw HTTP data. Note that data in the figure is truncated using "..." notation.}
    \label{fig:raw_http}
\end{figure}

\textbf{Raw and Standardized Telemetry Analysis:} During the manual analysis phase, our primary goal is to identify high-fidelity attacker signatures in each of the three phases of the attack chain. In this example, for Initial Access through Exploit Public-Facing Application, we will be looking for the initial Shellshock payload in the HTTP User-Agent field. This signature is shown in the raw HTTP request in Figure \ref{fig:raw_http}. Next, for the Execution via Exploitation for Client Execution techniques, we will look for either binary execution of the initial payload, downloading of a secondary payload, or execution of the secondary payload. Finally, for the Command and Control establishment, we will look for egress connections from the target to the attacker machine over port 4444.

\textbf{LLM-Assisted Signature Identification: }
To identify many of the low-fidelity signatures present in our attacks, we utilize the Claude Opus 4 model to scan and analyze our raw and standardized logs. The logs for SETC for each exploit will only contain telemetry from the target and attacker machines during the time of exploitation. The log sizes are small enough to fit into the context window of an LLM using a retrieval augmented generation (RAG) service. For this example, we loaded all of the logs related to CVE-2014-6271 into a Claude project and used a detailed prompt to generate a report around attack signatures.

The analysis identified 16 total attack signatures across all of the raw and standardized logs. These signatures were categorized into various buckets, including buckets defined in our attack chain: Initial Access, Execution, and C2. While some of the signatures overlap with our high-fidelity signatures identified in the previous manual phase, the majority of the signatures are considered low-fidelity and are most likely not sufficient on their own to identify active exploitation with high confidence. Examples of some of these low-fidelity signatures from the example LLM report include:
\begin{itemize}

    \item Large User-Agent lengths in HTTP requests
    \item Bash spawning processes from the Apache user on the target machine
    \item Execution of processes by the www-data user on the target machine
    \item Large data transfers
    \item Asymmetric data flows
    \item Parent-child process relationships
    \item Network connections being established and not terminated.
\end{itemize}

The LLM analysis and reports are then used to generate coverage gap statistics for each signature across all logging standards. Table \ref{tab:llm_example} presents the coverage analysis of the 12 signatures identified in the report, which relate to initial access, process execution, and C2 establishment.

\begin{table}[h!]
    \centering
    \caption{Signature coverage for Shellshock example}
    \begin{tabular}{|c|c|c|c|}\hline
         Log Source&  Initial Access&  Process Execution& C2\\\hline
         CIM& 100\%(3/3) & 100\%(4/4) & 40\%(2/5) \\\hline
         OCSF& 100\%(3/3) & 75\%(3/4) & 40\%(2/5) \\\hline
         ECS& 100\%(3/3) & 75\%(3/4) & 40\%(2/5) \\ \hline
    \end{tabular}
    \label{tab:llm_example}
\end{table}

\textbf{LLM Validation and Error Correction: } Manual review of the LLM-generated report identified several critical errors requiring correction. Most significantly, the LLM initially claimed Zeek had 100\% coverage for Execution via Exploitation for Client Execution technique signatures. In this case, the LLM failed to recognize that, as a network monitoring tool, Zeek cannot observe host-based process creation events. This error cascaded into miscalculated total coverage percentages and misleading claims about Zeek's ability to detect process execution. After correction, the revised LLM report accurately reflects that Zeek achieves 0\% coverage for execution signatures and maintains a high coverage for Initial Access and Command and Control signatures.

\textbf{Effectiveness \& Detection Score} 
The final step in analyzing all CVEs in this research is to calculate an effectiveness and detection score. The effectiveness score is a percentage score that measures the overlap of signatures found in the raw telemetry versus the log standards. This is an aggregation of the metrics shown in Table \ref{tab:llm_example}.

Detection scores are a binary measurement indicating the logging standard's ability to detect high-fidelity signatures across all three phases of our attack chain. In the manual phase 3 of this example, we determined that all three logging standards provided ample visibility across our attack chain. The final scores for CVE-2014-6271 are shown in Table \ref{tab:my_label}

\begin{table}[h!]
    \centering
    \caption{Metrics results for example CVE-2014-6271}
    \begin{tabular}{|c|c|c|}\hline
         Standard&  Effectiveness& Detection\\\hline
         CIM&  75\%& True\\\hline
         OCSF&  66\%& True\\\hline
         ECS&  66\%& True\\ \hline
    \end{tabular}
    \label{tab:my_label}
\end{table}

\subsection{Evaluation Results}

This section presents the comprehensive results of our evaluation of logging standard effectiveness for detecting remote exploitation of vulnerabilities defined in our attack chain. Through the analysis of 50 diverse RCE vulnerabilities using the SETC framework, we quantified both the telemetry preservation capabilities and practical detection efficacy of the three major logging standards in our study, CIM, OCSF, and ECS. Our evaluation reveals variations in how effectively each standard captures security-relevant information across the different phases of the attack chain, presenting findings relevant to researchers and practitioners using these logging standards. The following sections present our effectiveness scoring results, detection rate results, and key findings that emerged from this evaluation.

\subsubsection{Effectiveness Score Analysis}

The comprehensive results of the effectiveness scores reveal distinct security telemetry performance differences among the evaluated logging standards when measuring their ability to preserve security-relevant data against raw telemetry sources. As shown in Table \ref{tab:eff_score}, CIM demonstrated the highest overall effectiveness at 63\%, followed by OCSF at 58\%, and ECS at 57\%. This performance advantage was consistent across all three phases of our attack chain. For Exploit Public Application signatures, CIM captured 67\% compared to OCSF's 60\%, and ECS's 60\%. The gap widened further in Client Execution telemetry, where CIM preserved 73\% of signatures versus 70\% for OCSF, and 69\% for ECS. Command and Control signatures showed the most significant degradation across all standards, with CIM retaining only 50\%, OCSF 45\%, and ECS 43\%.

\begin{table}[h!]
    \centering
\caption{Log standard effectiveness scores based on low-fidelity LLM identified signatures}
\label{tab:eff_score}
    \begin{tabular}{|>{\centering\arraybackslash}p{0.06\linewidth}|>{\centering\arraybackslash}p{0.17\linewidth}|>{\centering\arraybackslash}p{0.17\linewidth}|>{\centering\arraybackslash}p{0.17\linewidth}|>{\centering\arraybackslash}p{0.17\linewidth}|}\hline
         &  Total Signatures&  Exploit Public Application&  Client Execution& Command \& Control\\\hline
         CIM&  63\% (468/743)&  67\% (161/242)&  73\% (185/254)& 50\% (122/246)\\\hline
         OCSF&  58\% (434/743)&  60\% (145/242)&  70\% (179/254)& 45\% (110/246)\\\hline
         ECS&  57\% (427/743)&  60\% (146/242)&  69\% (174/254)& 43\% (107/246)\\ \hline
    \end{tabular}

\end{table}

The effectiveness analysis also revealed interesting patterns when comparing web-based versus service-based vulnerabilities, as detailed in Table \ref{tab:eff_score_vs}. For web vulnerabilities, comprising 80\% of our corpus, CIM maintained its lead with 64\% effectiveness (377/589), while OCSF and ECS both achieved 58\% (343/589 and 342/589, respectively). The performance differential narrowed noticeably for service-based vulnerabilities. For service-based vulnerabilities, CIM and OCSF both achieved 59\% effectiveness (91/153) while ECS dropped to 56\% (85/153). This difference indicates that CIM's advantage in this study primarily stems from a more comprehensive set of required fields in its web-based telemetry, particularly in capturing HTTP-specific data. The consistent performance degradation in Command and Control telemetry across all standards indicates a systemic limitation in how current logging standards record outbound connection metadata and process execution context.

\begin{table}[h!]
    \centering
\caption{Log standard effectiveness scores for web (HTTP) vs service (non-HTTP) based vulnerabilities for low-fidelity LLM identified signatures}
\label{tab:eff_score_vs}
    \begin{tabular}{|>{\centering\arraybackslash}p{0.06\linewidth}|>{\centering\arraybackslash}p{0.17\linewidth}|>{\centering\arraybackslash}p{0.17\linewidth}|>{\centering\arraybackslash}p{0.17\linewidth}|>{\centering\arraybackslash}p{0.17\linewidth}|}\hline
         &  Total Signatures&  Exploit Public Application&  Client Execution& Command \& Control\\\hline
         Web Vuln CIM&  64\% (377/589)&  70\% (133/191)&  74\% (146/197)& 49\% (98/201)\\\hline
         Web Vuln OCSF&  58\% (343/589)&  61\% (116/191)&  71\% (139/197)& 44\% (88/201)\\\hline
         Web Vuln ECS&  58\% (342/589)&  63\% (120/191)&  70\% (137/197)& 42\% (85/201)\\\hline 
         Service Vuln CIM&  59\% (91/153)&  55\% (28/51)&  68\% (39/57)& 53\% (24/45)\\\hline
         Service Vuln OCSF&  59\% (91/153)&  57\% (29/51)&  70\% (40/57)& 49\% (22/45)\\\hline
         Service Vuln ECS&  56\% (85/153)&  51\% (26/51)&  65\% (37/57)& 49\% (22/45)\\ \hline 
    \end{tabular}

\end{table}

\subsubsection{Detection Score Analysis}

While effectiveness scores measure the granular preservation of signatures, the detection score analysis reveals a more nuanced picture of practical exploit detection capabilities. As shown in Table \ref{tab:det_score}, all three logging standards demonstrated identical detection rates across our vulnerability corpus. Initial Access detection through Exploit Public Application signatures was detected in only 26\% (13/50) of cases. In comparison, Client Execution detection improved dramatically to 76\% (38/50), and Command and Control establishment was detectable in 90\% (45/50) of vulnerabilities. All standards achieved only 20\% (10/50) full detection coverage when requiring high-fidelity signatures across all attack phases. This low full detection rate indicates significant variation in phase-specific detection capabilities. This finding stems from gaps in logging standard HTTP and network logs and is presented in greater detail in the findings section of this paper.

\begin{table}[h!]
    \centering
\caption{Log standard detection scores based on high-fidelity manually identified signatures}
\label{tab:det_score}
    \begin{tabular}{|>{\centering\arraybackslash}p{0.07\linewidth}|>{\centering\arraybackslash}p{0.17\linewidth}|>{\centering\arraybackslash}p{0.17\linewidth}|>{\centering\arraybackslash}p{0.17\linewidth}|>{\centering\arraybackslash}p{0.17\linewidth}|}\hline
         &  Full Detection&  Exploit Public Application&  Client Execution& Command \& Control\\\hline
         CIM&  20\% (10/50)&  26\% (13/50)&  76\% (38/50)&  90\% (45/50)\\\hline
         OCSF&  20\% (10/50)&  26\% (13/50)&  76\% (38/50)& 90\% (45/50)\\\hline
         ECS&  20\% (10/50)&  26\% (13/50)&  76\% (38/50)& 90\% (45/50)\\ \hline
    \end{tabular}

\end{table}

The detection analysis exposed a critical distinction between web and service-based vulnerabilities, as detailed in Table \ref{tab:http_det_score}. Web-based vulnerabilities demonstrated marginally better detection rates, with 25\% (10/40) achieving complete kill chain visibility, and Initial Access detection at 33\% (13/40). Notably, service-based vulnerabilities demonstrated 0\% detection for both the whole kill chain and Initial Access phases, despite maintaining comparable detection rates for Client Execution (70\%) and Command and Control (90\%). This complete failure to detect Initial Access for service vulnerabilities highlights a fundamental limitation in how logging standards capture non-standard network application data. 

\begin{table}[h!]
    \centering
\caption{Log standard detection scores for web (HTTP) vs service (non-HTTP) based vulnerabilities}
\label{tab:http_det_score}
    \begin{tabular}{|>{\centering\arraybackslash}p{0.07\linewidth}|>{\centering\arraybackslash}p{0.17\linewidth}|>{\centering\arraybackslash}p{0.17\linewidth}|>{\centering\arraybackslash}p{0.17\linewidth}|>{\centering\arraybackslash}p{0.17\linewidth}|}\hline
         &  Full Detection&  Exploit Public Application&  Client Execution& Command \& Control\\\hline
         Comb-ined& 20\% (10/50)&  26\% (13/50)&  76\% (38/50)&  90\% (45/50)\\\hline
         Web& 25\% (10/40)&  33\% (13/40)&  78\% (31/40)&  90\% (36/40)\\\hline
         Service& 0\% (0/10)&  0\% (0/10)&  70\% (7/10)&  90\% (9/10)\\\hline
    \end{tabular}
\end{table}

\subsubsection{Key Findings}
Our comprehensive evaluation of logging standards revealed several critical insights that have significant implications for security monitoring and logging standard development. These findings emerged from analyzing both quantitative metrics and qualitative patterns across our vulnerability corpus, providing evidence-based guidance for security practitioners and standard developers alike.

\textbf{This research successfully established a valid and reproducible measurement methodology for comparing logging standards in a security context.} Through the combination of effectiveness scores and detection rates, we created a quantitative framework that objectively evaluates logging standard performance across diverse vulnerability types and attack phases. The methodology's strength lies in its dual-metric approach: effectiveness scores provide granular measurement of signature preservation from raw telemetry to standardized formats, while detection scores assess practical security monitoring capabilities. This measurement framework proved capable of revealing both subtle performance differences between standards and systemic gaps across all evaluated frameworks. The reproducible nature of SETC-based evaluation ensures that these measurements can be validated, extended, and applied to future logging standards or expanded vulnerability population sets, establishing a foundation for evidence-based logging standard selection and improvement.

\textbf{Focusing on only required fields, CIM emerged as the most effective logging standard across all evaluation metrics.} The Common Information Model consistently outperformed both OCSF and ECS in preserving security-relevant telemetry. With our corpus data, CIM achieved an overall effectiveness score of 63\% compared to 58\% for OCSF and 57\% for ECS. This performance advantage was most evident in web-based vulnerabilities, where CIM's broader required HTTP field definitions captured 6\% more signatures than competing standards. CIM's superiority here stems from its more expansive set of required fields. This expanded set of required fields helps the standard perform better in capturing fields related to low-fidelity signatures, such as packet sizes and categories. While all three standards can achieve feature parity through the use of optional fields or custom fields, CIM's richer required fields provide security teams with more guidance in logging contextual data for threat hunting, forensic analysis, and low-fidelity indicator correlation. 

\textbf{Process logs emerged as the most valuable source of telemetry for detecting exploitation activity.} Our analysis consistently demonstrated that process telemetry provided the highest-fidelity indicators across all vulnerability types and logging standards. With 76\% detection rates for Client Execution and 90\% for Command and Control establishment, process logs captured critical evidence of successful exploitation, including command execution, payload deployment, and persistence mechanisms. Process logs uniquely preserved the execution context, parent-child relationships, and command-line arguments, which directly revealed the activities of the attacker. This telemetry source proved especially valuable because it captured post-exploitation behavior regardless of the initial access vector, providing a reliable detection point even when network and web logs failed to capture exploitation signatures.

In our study, process logs also provided the highest signal-to-noise ratio among all log sources. Shown in Table \ref{tab:ls_volumes}, mean process log volume was 15.3 events per exploit (with a standard deviation of 19). In contrast, HTTP and network logs, despite their higher volumes (means of 20 and 29 events respectively, with standard deviations of 105 each), contained more ancillary data that, while useful for correlation and investigation, proved less definitive for confirming successful exploitation.

\begin{table}[h!]
    \centering
\caption{Log standard event volumes for each exploit}
\label{tab:ls_volumes}
    \begin{tabular}{|>{\centering\arraybackslash}p{0.13\linewidth}|>{\centering\arraybackslash}p{0.15\linewidth}|>{\centering\arraybackslash}p{0.15\linewidth}|>{\centering\arraybackslash}p{0.16\linewidth}|>{\centering\arraybackslash}p{0.16\linewidth}|}\hline
         &  Min&  Max&  Mean& Stdev\\\hline
         HTTP&  0&  732&  20& 105\\\hline
         Network&  5&  740&  29& 105\\\hline
         Process&  2&  75&  15.3& 19\\\hline
         Combined&  9&  1474&  64& 209\\ \hline
    \end{tabular}

\end{table}

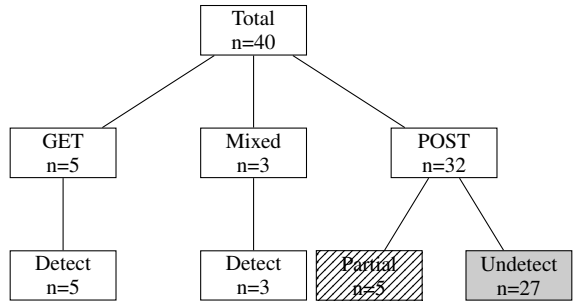
\begin{figure}[h!]
\centering
\begin{tikzpicture}[
    level 1/.style={sibling distance=2.8cm, level distance=1.8cm},
    level 2/.style={sibling distance=2.2cm, level distance=1.8cm},
    every node/.style={rectangle, draw=black, align=center, font=\footnotesize, minimum width=1.4cm, inner sep=2pt},
    detectable/.style={fill=white},
    undetectable/.style={fill=black!20},
    partial/.style={pattern=north east lines},
    scale=0.9
]

\node {Total\\n=40}
    child {node {GET\\n=5}
        child {node[detectable] {Detect\\n=5}}
    }
    child {node {Mixed\\n=3}
        child {node[detectable] {Detect\\n=3}}
    }
    child {node {POST\\n=32}
        child {node[partial] {Partial\\n=5}}
        child {node[undetectable] {Undetect\\n=27}}
    };

\end{tikzpicture}
\caption{Hierarchical breakdown of HTTP vulnerability detection capabilities. Shaded nodes indicate undetectable vulnerabilities.}
\label{fig:vulnerability-tree}
\end{figure}

\textbf{HTTP POST data represents a critical gap in current logging standards.} The evaluation revealed that 67.5\% (27/40) of web-based vulnerabilities in this study remained completely undetectable in our logging standards when exploits were delivered via POST request bodies. None of the evaluated logging standards include POST body data as a required or even recommended field, creating a massive blind spot for web application security monitoring. This gap is particularly concerning given that POST-based exploitation represents the predominant attack vector for modern web vulnerabilities. While GET-based exploits achieved 100\% detection due to payload visibility in URL parameters, the security community's privacy-conscious exclusion of POST data from logging standards has inadvertently eliminated visibility into the majority of web-based attacks. The hierarchical breakdown in Figure \ref{fig:vulnerability-tree} further illustrates this gap by showing, among the 40 web vulnerabilities, only GET-based exploits (5/5) and mixed GET/POST exploits (3/3) achieved reliable detection. In comparison, 27 of 32 POST-only vulnerabilities remained undetectable due to the absence of POST body data in standard log fields. The disparity highlights how current logging standards, developed with privacy and storage constraints in mind, may inadvertently favor the detection of post-exploitation activity over initial compromise attempts.

\textbf{The absence of raw network payload data creates similar critical gaps for service vulnerabilities.} Our evaluation demonstrated a complete detection failure (0\%) for Initial Access signatures across all non-HTTP service vulnerabilities. This gap stems from the focus of logging standards on metadata and connection summaries rather than packet contents. Unlike web logs that selectively capture some HTTP fields, network logs in standardized formats typically contain only flow-level information similar to NetFlow, omitting the actual protocol payloads where exploitation signatures reside. This limitation proved especially problematic for vulnerabilities in services such as SMB, SSH, and database protocols, where exploit signatures exist exclusively within protocol-specific packet data that logging standards do not preserve.

\section{Limitations}
Our research does not take into account stealthy attack approaches, as exploits selected to adhere to the RCE attack chain were designed to provide uniform and predictable tactics, techniques, and procedures. Following design-science principles, we deliberately constrained exploit diversity to enable systematic comparison across controlled conditions, isolating logging standard capabilities from confounding attack variations. All Metasploit payloads, where possible, were configured to deploy a remote access tool connecting to the command and control server over a reverse TCP connection on port 4444. All components of the attack chain relating to Client Execution and Command and Control establishment would likely have more variation in real-world environments. 

The research investigated only one attack chain, and while this attack chain is very prevalent and covers a broad scope of attack tactics, other attack chains may have increased or decreased logging standard metrics. Our focus on the Initial Access → Execution → Command and Control progression, while representative of common exploitation patterns, does not capture the full spectrum of adversarial behaviors. Attack chains involving lateral movement, privilege escalation, or persistence mechanisms might reveal different gaps or strengths in logging standards that are not reflected in our current evaluation.

Although the research uses a relatively large vulnerability and exploit population compared to typical academic research, 50 vulnerabilities are still a relatively small population representation compared to the number of CVEs released yearly. This sample size, while sufficient for establishing initial patterns and trends, may not capture the full diversity of vulnerability classes, exploitation techniques, or logging requirements present in the broader threat landscape. The development time and practical constraints of manually creating, integrating, configuring, and testing each vulnerability within the SETC framework limited our ability to scale to larger population sizes.

Our evaluation is also constrained by the current capabilities of the SETC framework, limiting our analysis to Linux-based systems and Metasploit-compatible exploits. However, these represent deliberate design tradeoffs. The Metasploit dependency provides standardized exploit modules that ensure experimental reproducibility and systematic comparison. Additionally, SETC's plugin-based architecture supports future expansion to additional operating systems, payload types, and attack frameworks without requiring fundamental redesign. This constraint currently excludes vulnerabilities affecting Windows environments, mobile platforms, and cloud-native services, which may exhibit different logging patterns and detection requirements. The requirement for containerized vulnerable services may not fully represent the complexity of production environments where logging standards are typically deployed.

Finally, the controlled container environments used in this research lack the baseline telemetry, which is often collected in production environments. This limitation particularly affects our ability to evaluate anomaly-based detection capabilities that depend on establishing normal behavior patterns. In real-world deployments, logs from logging standards can be used to distinguish between legitimate activities and malicious exploitation by comparing baseline data against new activity. This detection approach is not fully captured in our isolated test environment. Our focus on descriptive signature metrics is analytically justified for foundational research. By focusing analysis on attack chain signatures in standardized logs, we establish the empirical foundation for understanding logging capabilities. This foundation can then be built upon to analyze detection strategies.

\section{Related Work}
Few to no academic papers focus on the analysis, use, or effectiveness of logging standards. However, various academic papers have investigated and analyzed the effectiveness of telemetry and storage techniques in enhancing detection tools or capabilities. Two notable publications propose detection systems based on different telemetry sources to identify security events through graph analysis. 

RapSheet\cite{rapsheet} provides an approach through tracing initial infection points (IIP) and associated endpoint events with tactical provenance graphs (TPG) to better identify attackers with endpoint detection and response (EDR) telemetry. For the initial research, the approach focuses on MITRE ATT\&CK techniques and chained attack templates, referred to as skeleton graphs, to increase fidelity in their detections. The proposed system, RapSheet, also offers a novel approach to reducing EDR log volumes.

SHADEWATCHER\cite{shade} proposes using provenance graphs to identify attack activity through system calls. The proposed system SHADEWATCHER was created to identify attack chains through graph entity correlations. Similar to the system Rapsheet, the groups of defined provenance graph node correlations can be used to identify attacker activity in voluminous event sources. 

As this work has significant dependencies on SETC, it is relevant to review related literature focusing on the automated deployment of cyber ranges and security test environments. Labtainers\cite{labtainers1}\cite{labtainers2} is a system that creates monitored security environments. While its intended purpose mainly involves deploying security exercises for educational purposes, its auditing and monitoring capacities demonstrate approaches relevant to telemetry collection frameworks like SETC. While cyber ranges are not designed to deploy end-to-end automated system compromise techniques, they are related to deploying preconfigured environments for security exercises. Two notable cyber range deployment frameworks include Crack\cite{crack} and CRATE\cite{crate}. These systems focus on environment deployment capabilities, and their approaches differ significantly from SETC's end-to-end exploitation and telemetry collection focus. 

There are a limited number of security test environments similar to SETC. However, they do exist. TestREx \cite{textrex} is one of the most comparable works to the project. The system is intended to provide tooling to systematically test exploits against web applications in a controlled and reproducible environment. However, the architecture and container framework are designed to run vulnerabilities and exploits all in a single container. This type of design limits the telemetry, attack realism, and container reuse provided by SETC. Another similar system is BugBox \cite{bugbox}. While BugBox provides a framework that can perform automated end-to-end vulnerability testing, it is explicitly built for deploying and testing PHP applications.

Finally, there is sometimes a misconception and confusion that SETC has overlapping features with malware sandboxes. Malware sandboxes such as CucKoo Sandbox\cite{cukoo}, Triage, and VirusTotal focus on detonating, analyzing, and testing malware. While SETC could test and analyze malware similar to these systems, that is not its intended use case. SETC and this research focus on deploying vulnerable systems, exploiting these vulnerabilities, and recording telemetry of the attack. Malware sandboxes have limited capability for this type of analysis and typically focus on a malware sample running on a single system.

\section{Future Work}
Our evaluation of logging standard effectiveness in this research reveals several promising avenues for future research. These avenues could significantly enhance the scope and depth of the presented telemetry analysis and other research paths. The avenues build upon the foundation established by SETC, and our findings presented in this paper. They address existing limitations and also explore emerging technologies.

\subsection{Dynamic Logging Standards}
One exciting area for future development involves dynamically extending logging standards through machine learning approaches similar to Log2Vec. While fully dynamic log template creation approaches lack the core benefits of established logging standards due to their entirely dynamic nature, they present significant opportunities when used in conjunction with existing standards. Using them to expand the capabilities of existing standards would allow them to gain the benefits of both approaches.

Dynamic field generation used in optional or extended fields would address several key limitations observed in our evaluation. First, it could provide automated gap coverage by identifying and creating fields for exploit signatures that existing standards fail to capture. Second, it could enhance security visibility by dynamically adapting to logging sources, attack techniques, and vulnerability classes. Finally, it could enable the safe propagation of data through intelligent field selection within logs or log elements typically excluded due to their tendency to contain sensitive data. The potential to create self-adapting logging schemas that evolve with the threat landscape while maintaining the benefits of established standards represents a significant advancement in security telemetry collection.

\subsection{Offensive Security AI Agents}
Past research efforts with SETC investigated the possibility of using automated offensive security agents\cite{iapts}\cite{vape}\cite{offsecautomation} to provide larger vulnerability corpuses and more realistic exploitation scenarios. However, traditional automated exploitation tools lacked the complexity and capability to enhance our research significantly. These early approaches were primarily limited to pre-programmed attack patterns, brute force methods, or text-matching decision algorithms that were prone to error.

Recent advances in offensive security agents powered by artificial intelligence and large language models present far more promising opportunities\cite{singer2025feasibilityusingllmsautonomously}\cite{deng2024pentestgpt}\cite{shao2024nyu}\cite{zhu2024teams}. These AI-driven agents demonstrate capabilities that could dramatically expand both the attack techniques and vulnerability corpuses available for our research. Unlike their predecessors, modern AI agents show promise in their ability to adapt exploitation strategies, reason about attack vectors, and potentially discover new exploitation techniques during their operation.

Future research investigating how these AI agents perform within the SETC framework could greatly expand the scope of our logging standard evaluation. AI agents could automate the time-intensive process of vulnerability corpus creation and generate more diverse attack scenarios. This automation could enable the assessment of logging standards against a significantly larger set of vulnerabilities than those presented in our current research.

Additionally, SETC could serve as a valuable platform for measuring the effectiveness and capabilities of these emerging offensive security AI agents \cite{cybergym}. SETC can be used to provide a controlled, reproducible environment for AI agent operation. This controlled environment could contribute to understanding the capabilities, limitations, detection patterns, and potential risks associated with AI-powered offensive security tools. This bidirectional research relationship could advance both logging standard evaluation methodologies and our understanding of true capabilities of offensive security AI agents. 

\subsection{Dynamic Honeypots}
The creation and integration of specialized honeypots represents another promising research direction that could significantly expand the population sets available for our research. Traditional honeypots\cite{honeypot} require HTTP or network service endpoints to be statically designed for specific vulnerabilities or attack types. This static design limits their utility for broad-scale logging standard evaluation and presents another manual bottleneck for our corpus creation.

The development of specialized honeypots capable of dynamically accepting a wide range of exploitation attempts could dramatically increase our research populations\cite{dynamichoneypot}. These adaptive honeypots would be designed to simulate vulnerable services across multiple vulnerability classes while capturing detailed telemetry about exploitation attempts. Unlike static vulnerable services, dynamic honeypots could provide sufficient exploit log telemetry to extend our research.

\section{Conclusion}
This research presents the first systematic, empirical evaluation of modern security logging standards using a novel methodology that combines design science and applied experimentation. We developed and validated a quantitative measurement framework that enables objective comparison of logging standard effectiveness in capturing security-relevant telemetry during active vulnerability exploitation. Through systematic analysis of 50 remote code execution vulnerabilities using the SETC framework, we established both granular effectiveness metrics and practical detection capabilities across CIM, OCSF, and ECS logging standards. Our dual-metric approach provided insights into logging standard telemetry coverage and practical effectiveness. Most importantly, it provided a reproducible methodology that can be validated, extended, and applied to future logging standards.

The evaluation of our corpus and defined attack chain identified several fundamental gaps in current logging standards that significantly impact security monitoring capabilities. Most critically, the absence of HTTP POST body data in all evaluated standards created a massive blind spot, rendering 67.5\% of web-based vulnerabilities completely undetectable when exploits were delivered via POST requests. Similarly, focusing on connection metadata rather than packet contents in network logs resulted in a complete failure to detect service-based vulnerabilities. These gaps stem from legitimate privacy and storage concerns but inadvertently eliminate visibility into predominant attack vectors.

In our data population, process telemetry emerged as the most valuable source for detecting exploitation activity, providing the highest-fidelity indicators across all vulnerability types and logging standards. With detection rates of 76\% for execution activities and 90\% for command and control establishment, process logs captured critical evidence of successful exploitation while maintaining superior signal-to-noise ratios compared to network and web logs. This finding underscores the importance of comprehensive process logging in security monitoring architectures and highlights where organizations should focus their telemetry collection efforts.

Along with our presented metrics, this research establishes a foundation for evidence-based selection and improvement of logging standards. Security practitioners can now make informed decisions about logging standard adoption based on empirical data rather than vendor claims or community momentum. Log standard developers gain insight into specific areas requiring enhancement to improve security telemetry coverage. The methodology and framework developed through this research provide a repeatable, scalable approach for continuous evaluation as both logging standards and threat landscapes evolve.

Our evaluation demonstrates that current logging standards capture only a fraction of available security telemetry when limited to required fields. This presents a critical decision organizations must make when implementing these standards in production environments. Security teams relying solely on required fields risk missing exploitation indicators for the majority of web-based attacks, while those implementing optional or custom fields gain significantly enhanced detection capabilities. This research provides quantitative evidence to guide these implementation decisions, enabling organizations to identify where additional logging investments yield maximum detection value. Most importantly, the measurement framework established here offers a path forward for continuous improvement of logging standards, ensuring they evolve alongside emerging threats while maintaining the interoperability and standardization benefits that make them valuable for enterprise security operations.

\section{Availability}\label{avail}
The SETC framework and data used for this research are fully open-source. SETC is available under the MIT license and can be found at \url{https://github.com/hackgnar/setc}.  Data sets and SETC configuration files used for this research can be found in the SETC open data sets repository found at \url{https://github.com/hackgnar/setc-data}.

\bibliographystyle{IEEEtranDOIandURLwithDate} 
\bibliography{ieee2024} 

\clearpage

\appendices
\onecolumn
\section{Vulnerability \& Exploit Corpus}\label{AppA}

\small
\begin{longtable}{|>{\RaggedRight\arraybackslash\hspace{0pt}}p{2.5cm}|>{\RaggedRight\arraybackslash\hspace{0pt}}p{4cm}|>{\RaggedRight\arraybackslash\hspace{0pt}}p{1.5cm}|>{\RaggedRight\arraybackslash\hspace{0pt}}p{8cm}|}
\caption{CVE Vulnerabilities and Associated Exploits} \label{tab:cve_vulnerabilities} \\
\hline
\multicolumn{1}{|c|}{\textbf{CVE}} & \multicolumn{1}{c|}{\textbf{Description}} & \multicolumn{1}{c|}{\textbf{Service}} & \multicolumn{1}{c|}{\textbf{MSF Exploit Module}} \\
\hline
\endfirsthead

\multicolumn{4}{c}%
{{\bfseries Table \thetable\ continued from previous page}} \\
\hline
\multicolumn{1}{|c|}{\textbf{CVE}} & \multicolumn{1}{c|}{\textbf{Description}} & \multicolumn{1}{c|}{\textbf{Service}} & \multicolumn{1}{c|}{\textbf{MSF Exploit Module}} \\
\hline
\endhead

\hline \multicolumn{4}{|r|}{{Continued on next page}} \\
\endfoot

\hline
\endlastfoot

CVE-2024-38856 & Apache OFBiz forgotPassword/ProgramExport RCE & HTTP & multi/http/apache\_ofbiz\_forgot\_password\_directory\_traversal \\
\hline
CVE-2024-36401 & Geoserver unauthenticated Remote Code Execution & HTTP & multi/http/geoserver\_unauth\_rce\_cve\_2024\_36401 \\
\hline
CVE-2024-27348 & Apache HugeGraph Gremlin RCE & HTTP & linux/http/apache\_hugegraph\_gremlin\_rce \\
\hline
CVE-2024-2044 & pgAdmin Session Deserialization RCE & HTTP & multi/http/pgadmin\_session\_deserialization \\
\hline
CVE-2023-46604 & ActiveMQ deserialization RCE & ActiveMQ & multi/misc/apache\_activemq\_rce\_cve\_2023\_46604 \\
\hline
CVE-2023-38646 & Metabase Setup Token RCE & HTTP & linux/http/metabase\_setup\_token\_rce \\
\hline
CVE-2023-37941 & Apache Superset Signed Cookie RCE & HTTP & linux/http/apache\_superset\_cookie\_sig\_rce \\
\hline
CVE-2023-32315 & Openfire authentication bypass with RCE plugin & HTTP & multi/http/openfire\_auth\_bypass\_rce\_cve\_2023\_32315 \\
\hline
CVE-2023-25194 & Apache Druid JNDI injection RCE & HTTP & multi/http/apache\_druid\_cve\_2023\_25194 \\
\hline
CVE-2023-21839 & Oracle WebLogic PreAuth Remote Command Execution via ForeignOpaqueReference IIOP Deserialization & HTTP & multi/iiop/cve\_2023\_21839\_weblogic\_rce \\
\hline
CVE-2022-46169 & Cacti unauthenticated command injection & HTTP & linux/http/cacti\_unauthenticated\_cmd\_injection \\
\hline
CVE-2022-22963 & Apache Spark Unauthenticated Command Execution & HTTP & linux/http/spark\_unauth\_rce \\
\hline
CVE-2022-22963 (b) & Spring Cloud Function SpEL Injection & HTTP & multi/http/spring\_cloud\_function\_spel\_injection \\
\hline
CVE-2022-22947 & Spring Cloud Gateway Remote Code Execution & HTTP & linux/http/spring\_cloud\_gateway\_rce \\
\hline
CVE-2022-0543 & Redis Lua RCE & Redis & exploit/linux/redis/redis\_debian\_sandbox\_escape \\
\hline
CVE-2021-42013 & HTTP apache normalize RCE & HTTP & exploit/multi/http/apache\_normalize\_path\_rce \\
\hline
CVE-2021-41773 & HTTP apache normalize RCE & HTTP & exploit/multi/http/apache\_normalize\_path\_rce \\
\hline
CVE-2021-25646 & Apache druid JS RCE & HTTP & exploit/linux/http/apache\_druid\_js\_rce \\
\hline
CVE-2021-25282 & SaltStack Salt API Unauthenticated RCE through wheel\_async client & HTTP & linux/http/saltstack\_salt\_wheel\_async\_rce \\
\hline
CVE-2021-22205 & GitLab Unauthenticated Remote ExifTool Command Injection & HTTP & multi/http/gitlab\_exif\_rce \\
\hline
CVE-2021-3129 & Laravel debug RCE & HTTP & exploit/multi/php/ignition\_laravel\_debug\_rce \\
\hline
CVE-2020-25646 & Apisix HTTP token RCE & HTTP & multi/http/apache\_apisix\_api\_default\_token\_rce \\
\hline
CVE-2020-17519 & Apache Flink JAR Upload Java Code Execution & HTTP & multi/http/apache\_flink\_jar\_upload\_exec \\
\hline
CVE-2020-16846 & SaltStack Salt REST API Arbitrary Command Execution & HTTP & linux/http/saltstack\_salt\_api\_cmd\_exec \\
\hline
CVE-2020-14882 & Oracle WebLogic Server Administration Console Handle RCE & HTTP & multi/http/weblogic\_admin\_handle\_rce \\
\hline
CVE-2020-11651 & SaltStack Salt Master/Minion Unauthenticated RCE & ZMQ & linux/misc/saltstack\_salt\_unauth\_rce \\
\hline
CVE-2020-9496 & Apache OFBiz XML-RPC Java Deserialization & HTTP & linux/http/apache\_ofbiz\_deserialization \\
\hline
CVE-2020-7247 & OpenSMTPD MAIL FROM Remote Code Execution & SMTP & unix/smtp/opensmtpd\_mail\_from\_rce \\
\hline
CVE-2019-17558 & Apache Solr Remote Code Execution via Velocity Template & HTTP & multi/http/solr\_velocity\_rce \\
\hline
CVE-2019-9082 & ThinkPHP Multiple PHP Injection RCE & HTTP & unix/webapp/thinkphp\_rce \\
\hline
CVE-2019-2725 & Oracle Weblogic Server Deserialization RCE & HTTP & multi/misc/weblogic\_deserialize\_asyncresponseservice \\
\hline
CVE-2018-11776 & Struts2 OGNL injection RCE & HTTP & multi/http/struts2\_multi\_eval\_ognl \\
\hline
CVE-2018-11770 & Apache Spark Unauthenticated Command Execution & HTTP & linux/http/spark\_unauth\_rce \\
\hline
CVE-2018-10933 & libssh Authentication Bypass & SSH & scanner/ssh/libssh\_auth\_bypass \\
\hline
CVE-2017-17562 & GoAhead Web Server LD\_PRELOAD Arbitrary Module Load & HTTP & linux/http/goahead\_ldpreload \\
\hline
CVE-2017-12636 & Apache CouchDB Arbitrary Command Execution & HTTP & linux/http/apache\_couchdb\_cmd\_exec \\
\hline
CVE-2017-12149 & JBoss invoker RCE & JBoss & exploit/multi/misc/jboss\_remoting\_unified\_invoker\_rce \\
\hline
CVE-2017-9841 & PHP Unauthenticated OS Command Execution & HTTP & unix/http/xdebug\_unauth\_exec \\
\hline
CVE-2017-7494 & Samba is\_known\_pipename() Arbitrary Module Load & Samba & linux/samba/is\_known\_pipename \\
\hline
CVE-2017-5638 & Apache Struts Jakarta Multipart Parser OGNL Injection & HTTP & multi/http/struts2\_content\_type\_ognl \\
\hline
CVE-2016-6811 & Hadoop YARN ResourceManager Unauthenticated Command Execution & HTTP & linux/http/hadoop\_unauth\_exec \\
\hline
CVE-2016-3088 & ActiveMQ exploit CVE-2016-3088 & HTTP & multi/http/apache\_activemq\_upload\_jsp \\
\hline
CVE-2016-3081 & Apache Struts Dynamic Method Invocation Remote Code Execution & HTTP & multi/http/struts\_dmi\_exec \\
\hline
CVE-2015-8562 & Joomla HTTP Header Unauthenticated Remote Code Execution & HTTP & multi/http/joomla\_http\_header\_rce \\
\hline
CVE-2014-6271 & Apache CGI Shellshock using user agent & HTTP & multi/http/apache\_mod\_cgi\_bash\_env\_exec \\
\hline
CVE-2012-1823 & PHP CGI Argument Injection & HTTP & multi/http/php\_cgi\_arg\_injection \\
\hline
CVE-2010-2075 & UnrealIRCD backdoor command execution & IRC & unix/irc/unreal\_ircd\_3281\_backdoor \\
\hline
CVE-2007-3280 & Postgres write to /tmp RCE & Postgres & linux/postgres/postgres\_payload \\
\hline
CVE-2007-2447 & Samba RCE & Samba & multi/samba/usermap\_script \\
\hline
CVE-2005-2877 & twiki history RCE & HTTP & exploit/unix/webapp/twiki\_history \\
\hline

\end{longtable}
\twocolumn

\end{document}